\documentstyle[11pt,paspconf,epsf]{article}

\begin{document}

\title{Multi-wavelength Observations of VHE BL Lac Objects}

\author{Michael Catanese}
\affil{Department of Physics and Astronomy, Iowa State University, Ames,
IA 50011-3160 \\ for the Whipple Collaboration}

\begin{abstract}

We present multi-wavelength observations of the very high energy (VHE,
E$\ga$250 GeV) $\gamma$-ray sources Mrk 421 and Mrk 501.  The VHE data
presented here were taken with the Whipple Observatory 10m
$\gamma$-ray telescope.  In the Mrk 421 campaign, conducted in 1995
April-May, correlations were observed between VHE $\gamma$-ray, X-ray,
extreme ultraviolet (EUV) and possibly R-band emission.  The X-rays
and $\gamma$-rays vary together with similar flux amplitude while the
EUV photons lag the X-rays and $\gamma$-rays by $\sim$1 day and have a
lower variability amplitude.  In the Mrk 501 campaign, conducted in
1997 April, a correlation was also observed between the VHE
$\gamma$-rays and X-rays with no delays apparent.  Here, the VHE
$\gamma$-ray variability is larger than the X-rays.  The observations
also indicate that the synchrotron spectrum extends to 100 keV before
cutting off.  For Mrk 421, the X-ray/VHE $\gamma$-ray correlation
indicates that $\delta \ga$38 and $B \ga 0.03$G.  For Mrk 501, the
correlation indicates a more modest $\delta \ga$2 and $B \ga$0.09G.

\end{abstract}


\section{Introduction}

At this time, three BL Lacertae objects (BL Lacs) have been detected
at very high energies (VHE, E$\ga$250 GeV): Markarian 421 (Mrk 421)
(Punch et al. 1992), Mrk 501 (Quinn et al. 1996), and 1ES 2344+514
(Catanese et al. 1998).  The VHE detections of Mrk 421 and Mrk 501
have been confirmed (e.g., Bradbury et al. 1997, Petry et al. 1996)
while the 1ES 2344+514 detection is still unconfirmed.  All three
objects are high frequency-peaked BL Lacs (HBLs) (Padovani \& Giommi
1995a), that is, the peak in the spectral energy distribution (SED),
plotted as $\nu F_\nu$, is in the X-ray region.  The VHE sources are
also the BL Lacs with the lowest known redshifts (Padovani \& Giommi
1995b): z=0.031 for Mrk 421, z=0.034 for Mrk 501, and z=0.044 for 1ES
2344+514.

Because of their variability at all wavelengths, BL Lacs can best be
understood through multi-wavelength observations.  The sensitivity of
VHE telescopes to sub-hour scale variability (cf., Gaidos et
al. 1996), their ability to detect low source 
fluxes, and the measurement of
spectra up to 10 TeV or more make the VHE observations an important
addition to multi-wavelength campaigns.  In particular, if the
$\gamma$-ray emission arises from inverse Compton scattering of the
same electron population which produces the low energy synchrotron
emission (e.g., Maraschi, Ghisellini, \& Celotti 1992, Dermer,
Schlickeiser, \& Mastichiadis 1992, Sikora, Begelman, \& Rees 1994,
Bloom \& Marscher 1996), the combination of VHE $\gamma$-ray and
synchrotron observations permit the estimation of the magnetic field
strength and Doppler factor in the jet where the $\gamma$-ray emission
is produced.

In this paper we describe a campaign on Mrk 421 conducted in 1995 (\S
2) and one on Mrk 501 completed in 1997 (\S 3).  Both campaigns gave
the first strong indication of correlated variability between VHE
$\gamma$-rays and lower energy emission in these objects.  Finally, we
compare the results of these observations and discuss their
implications (\S 4).

\section{1995 Markarian 421 campaign}

Between 1995 April 20 and May 5, Mrk 421 was monitored with the
Whipple Observatory $\gamma$-ray telescope (E$\sim$ 250 GeV -- 10
TeV), the Energetic Gamma-Ray Experiment Telescope (EGRET) on the {\it
Compton Gamma-ray Observatory} (E = 30 MeV -- 10 GeV), the {\it
Advanced Satellite for Cosmology and Astrophysics} ({\it ASCA}) (E =
0.7 -- 7.5 keV), the {\it Extreme Ultraviolet Explorer} ({\it EUVE})
($\lambda$ = 58 -- 174 \AA), and in the optical R band.  Less dense
monitoring of radio flux and optical polarization was also done.  The
results of these observations were first reported by Buckley et
al. (1996).  The light-curves for the contemporaneous observations are
shown in Figure~1.  EGRET observations resulted in an upper limit on
the flux above 100 MeV of 1.2$\times$10$^{-7}$ cm$^{-2}$ s$^{-1}$.

\begin{figure}[t!]
\plottwo{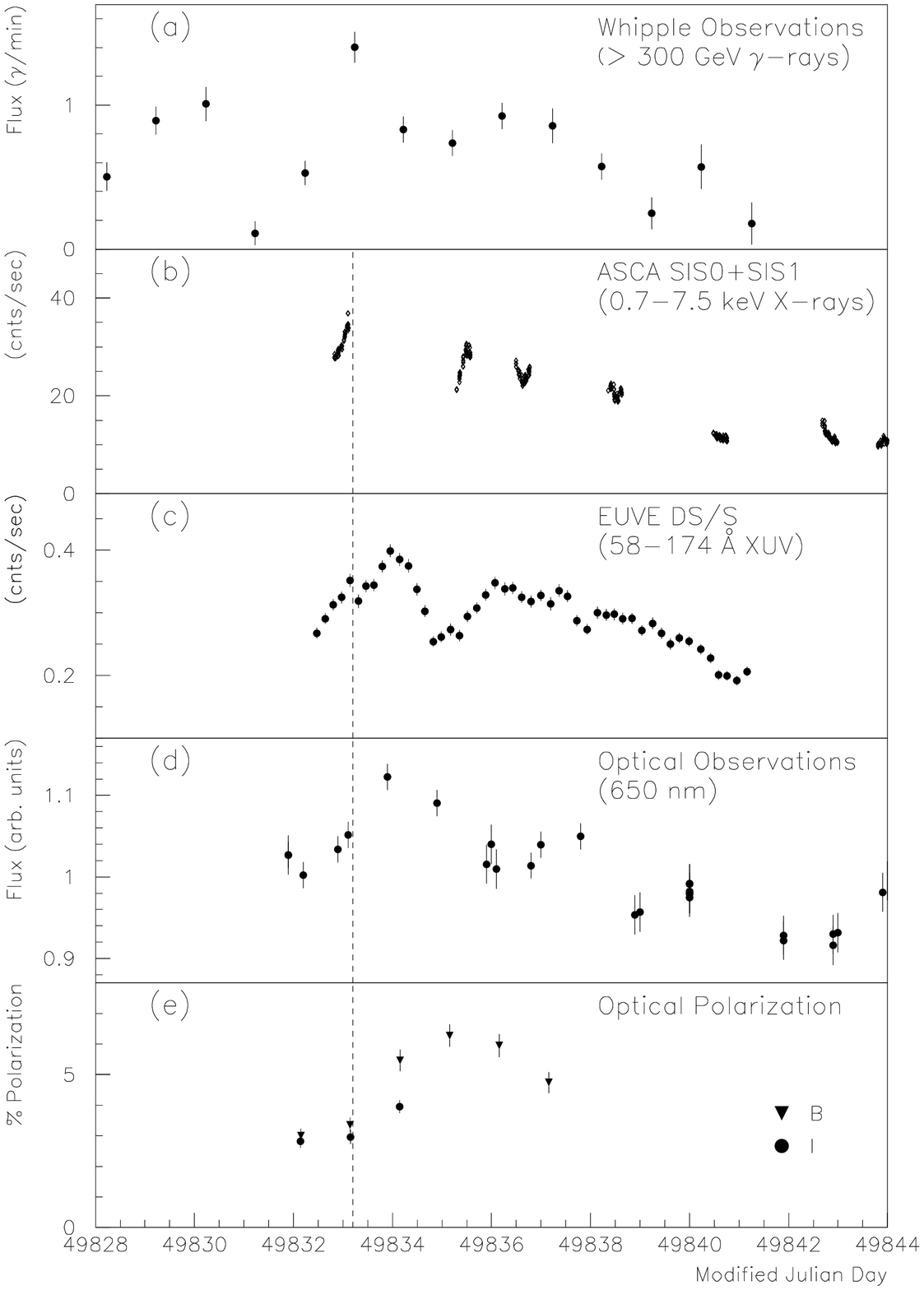}{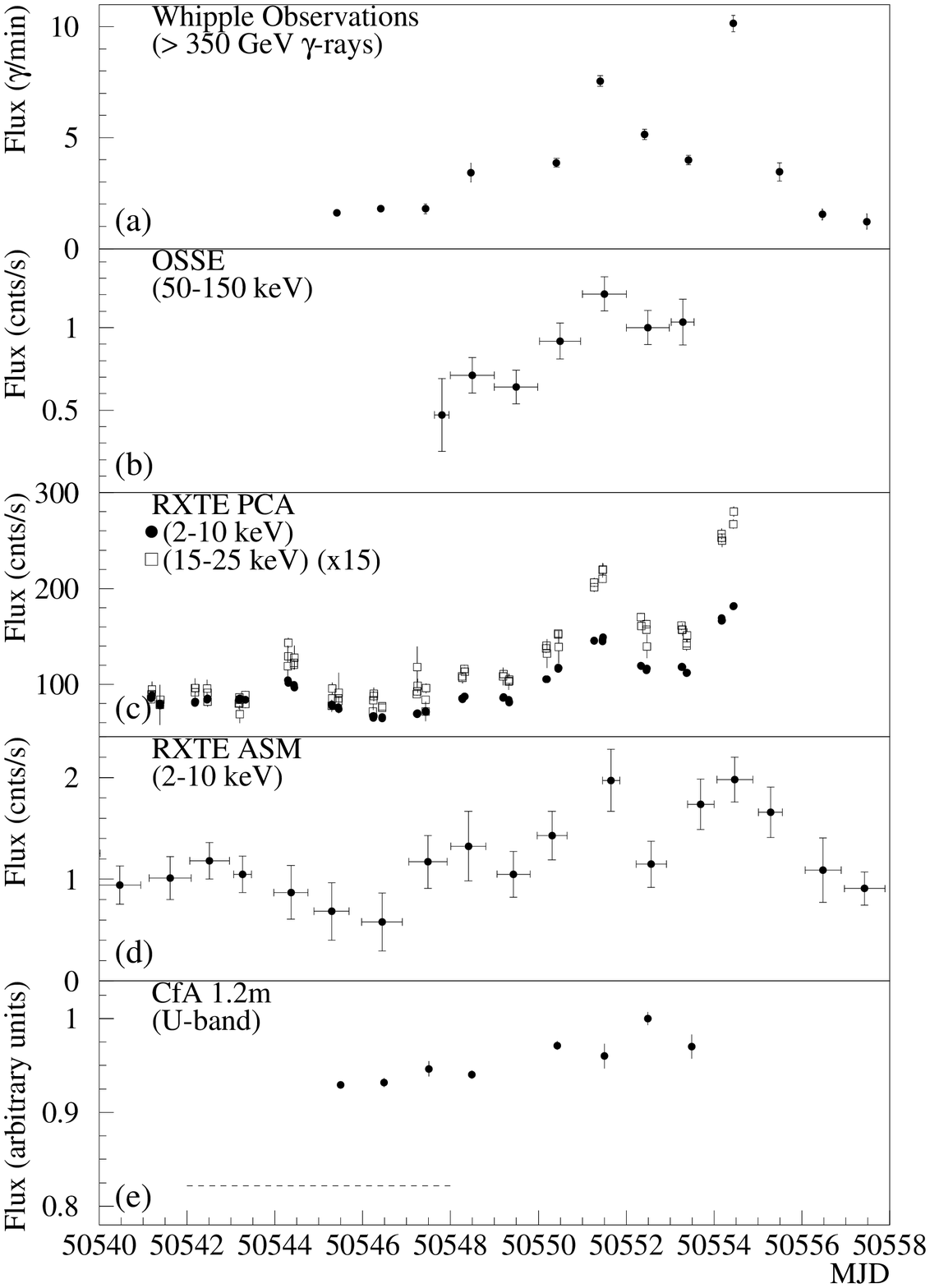}
\caption{{\it Left:} Multi-wavelength light-curve of Mrk 421 taken in
1995 April-May.  MJD 49833 corresponds to April 26.  Figure from
Buckley et al. (1996). {\it Right:} Contemporaneous light-curve of Mrk
501 taken in 1997 April.  MJD 50540 corresponds to April 2.  Figure
adapted from Catanese et al. (1997).}
\end{figure}

A prominent VHE flare is evident with peak flux on MJD 49833.  The VHE
flux is $\approx $0.75 times the VHE flux of the Crab Nebula.
Cross-correlation analysis of the VHE $\gamma$-rays and the X-rays
indicates a significant correlation with no time lag on day scales.
The amplitude of the flux variations of the VHE $\gamma$-rays and
X-rays are both $\approx$400\%.  A flare is also apparent in the {\it
EUVE} observations, and cross-correlation analysis indicates a maximum
correlation with a 1 day lag relative to the VHE $\gamma$-ray/X-ray
flare.  The amplitude of the {\it EUVE} flare was $\approx$200\%.  The
R-band and optical polarization observations also appear to correlate
with the VHE $\gamma$-ray/X-ray flare with a 1 day lag, but the
statistics of those observations do not permit a firm identification
of this variation with the X-ray/VHE $\gamma$-ray flare.  Indeed,
results reported by Wagner (1998) at this conference indicate that
optical observations conducted after the VHE $\gamma$-ray observations
had stopped show a flux increase while the X-ray flux continues to
decrease.  The amplitude of the R-band variation is only about 20\%
(after the galaxy light contribution is removed).  The U-band
polarization measurements show an $\approx$200\% variation in
amplitude.

\section{1997 Markarian 501 campaign}

Between 1997 April 2 and 20, Mrk 501 was observed with the Whipple
Observatory $\gamma$-ray telescope, EGRET, the Oriented Scintillation
Spectrometer Experiment (OSSE) (50 keV - 10 MeV), the {\it Rossi X-ray
Timing Experiment} ({\it RXTE}) (2 keV - 250 keV), and the Whipple
Observatory 1.2m optical telescope (UVBRI).  The results of the
Whipple Observatory, OSSE, and quicklook {\it RXTE} All-Sky Monitor
(ASM) observations were first reported by Catanese et al. (1997).
Here we add preliminary results from the Proportional Counter Array
(PCA) and High Energy X-ray Timing Experiment (HEXTE) on {\it RXTE}.
These latter observations consisted of two $\sim$1 ksec exposures per
night between April 2 and 15.  The light-curves of the contemporaneous
observations are shown in Figure~1.  For comparison, the count rate
from the Crab Nebula for the Whipple Observatory telescope during this
period was 2.7 $\gamma$/min.

During these observations, Mrk 501 exhibited two large amplitude,
day-scale VHE $\gamma$-ray flares (see Figure~1).  The VHE data are 
best fit by a spectrum which is parabolic when plotted as
$\log_{10}({\rm dN/dE})$ versus $\log_{10} {\rm E}$: ${\rm dN}/{\rm
dE} \propto {\rm E}^{-2.20 \pm 0.04 \pm 0.05 - (0.45 \pm
0.07)\log_{10}{\rm E}}$ (Samuelson et al. 1998).  The spectrum extends
to at least 8 TeV with no evidence of a cut-off and no spectral
variability is detected.  EGRET observations resulted in an upper
limit above 100 MeV of 3.7$\times$10$^{-7}$ cm$^{-2}$ s$^{-1}$ even
though the VHE flux was much higher during this period than it was
when EGRET detected Mrk 421.  In contrast, OSSE detected the highest
50 keV - 150 keV flux it had observed from any blazar (McNaron-Brown
et al. 1995) while it has never detected Mrk 421 in several
observations.  The spectrum measured by OSSE is fit well by a simple
power law with photon spectral index -2.08$\pm$0.15.  OSSE
observations show no evidence of spectral variability.  The PCA
spectra are best fit by broken power laws with a break consistently in
the range 5-6 keV.  The photon spectral indices vary from -1.9 to
-1.6 below the break energy and from -2.1 to -1.8 above the break
energy.  The average 2-10 keV flux during this period is $\approx 4.5
\times 10^{-10}$ erg cm$^{-2}$ s$^{-1}$.  

The correlation between the VHE $\gamma$-rays and the soft and hard
X-rays observed with {\it RXTE} and OSSE is clearly evident in Figure
1.  In the overlapping period of observations, April 9-15, a peak in
the flux is detected by all three instruments on April 13 (MJD 50551),
indicating time lags, if any, of $<$1 day.  The flux amplitude
variation between April 9 and April 13 is a factor of 4 for the VHE
$\gamma$-rays, 2.6 for 50-150 keV X-rays, 2.3 for 15-25 keV X-rays,
and 2.1 for 2-10 keV X-rays.  Optical measurements in the U-band
(without galaxy light subtraction) exhibit no simple correlation, but
the emission is significantly higher than in the previous month
(indicated by the dashed line in Figure 1).

\section{Discussion}

The multi-wavelength observations presented here show that Mrk 421 and
Mrk 501 exhibit a combination of similarities and differences.  First,
as indicated by their SEDs (see Figure 2), the high energy emission
does not dominate the power output of Mrk 421 and Mrk 501 as in many
of the EGRET blazars.  This is consistent with the expectations of
recent unification models for blazars because the HBLs have relatively
low luminosity (Fossati et al. 1998, Georganapoulos \& Marscher 1998).
But, while the power output for Mrk 421 in VHE $\gamma$-rays $\approx$
the power output in X-rays in several campaigns, for Mrk 501, the
power output in VHE $\gamma$-rays is only comparable to that in X-rays
during its highest emission states.  In the lower emission states, the
VHE $\gamma$-ray power output is somewhat lower.  In addition, though
the SED of Mrk 421 is typical of an HBL (peak synchrotron power output
at UV/soft X-ray energies with a trough in the power output in hard
X-rays where OSSE operates), Mrk 501 appears to be an extreme version
of an HBL - the synchrotron spectrum observed in 1997 peaks at
$\approx$100 keV before dropping off.  This is the highest extent of
the synchrotron spectrum ever seen in a blazar.  Also, the high energy
spectrum in Mrk 501 appears to be shifted to higher energies than in
Mrk 421, with a trough in the MeV-GeV range where EGRET operates.

\begin{figure}[t!]
\vspace*{-0.5in}
\plottwo{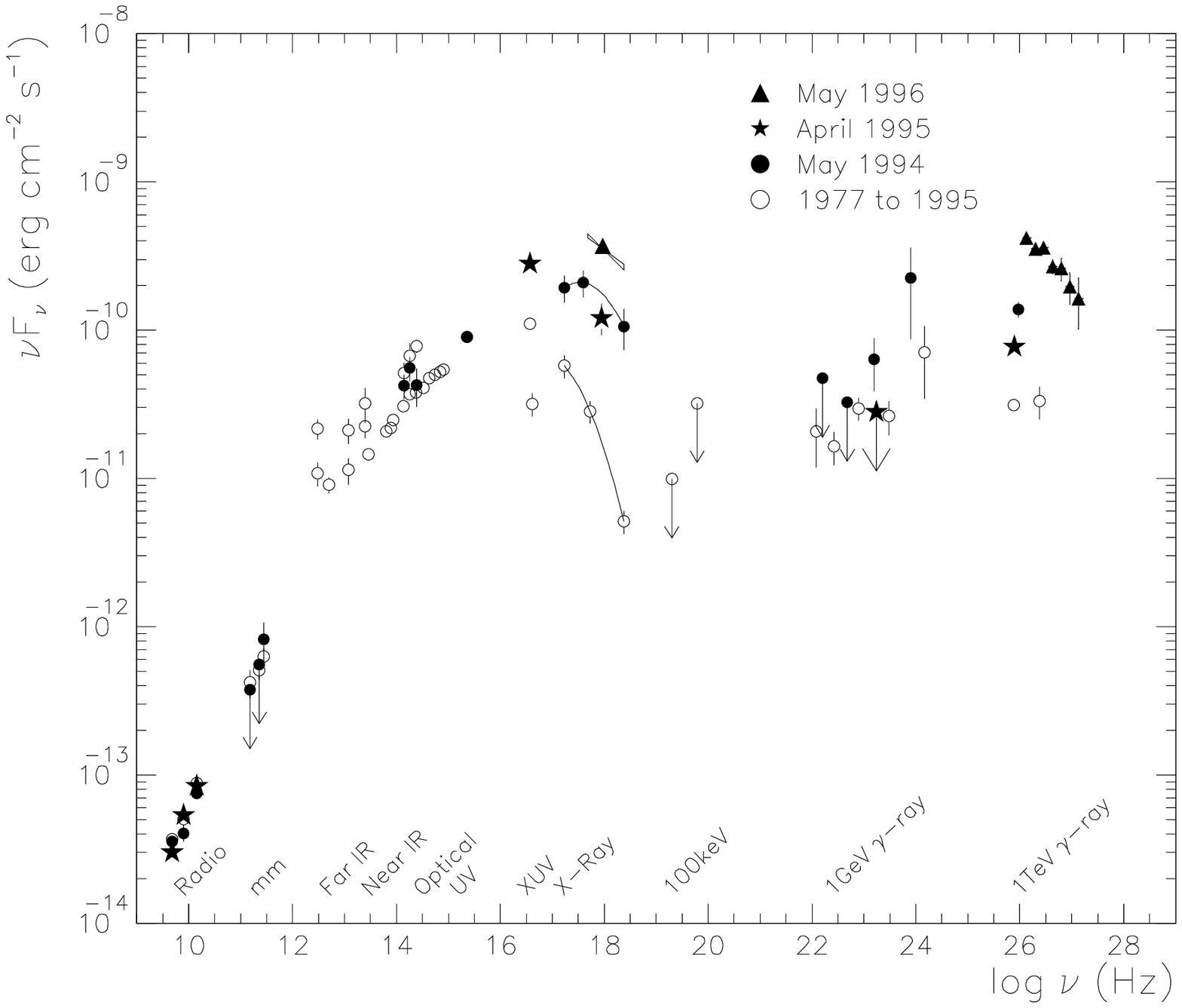}{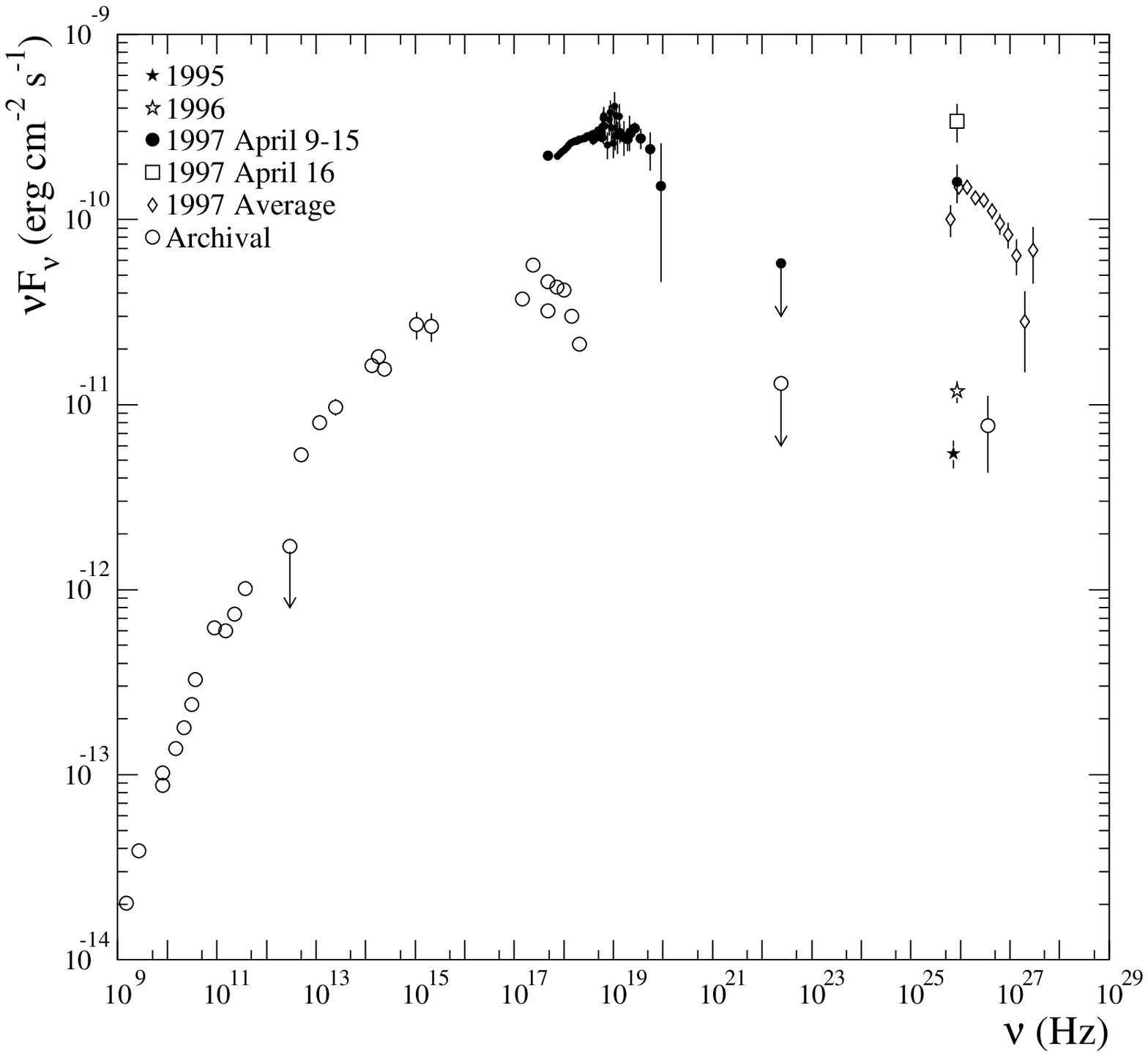}
\vspace*{-0.65in}
\caption{{\it Left:} Spectral energy distribution for Mrk 421.  Data 
presented here are indicated by the filled stars.  Figure from Buckley
et al. (1997).  {\it Right:} Spectral energy distribution for Mrk 501.
Data presented here are indicated by the filled circles.  Figure
adapted from Catanese et al. (1997) and Samuelson et al. (1998).}
\end{figure}

Second, for both Mrk 421 and Mrk 501, the relative variability
amplitude in the synchrotron band decreases with decreasing energy and
the VHE $\gamma$-rays and X-rays appear to be correlated with no
delays evident on day-scales.  This is consistent with flares produced
by impulsive increases in the efficiency for accelerating the highest
energy electrons (Marscher 1995, Mastichiadis \& Kirk 1997).  However,
the relative variability amplitude between the synchrotron and high
energy bands is quite different.  Mrk 421 exhibits similar variability
amplitude between VHE $\gamma$-rays and X-rays but in Mrk 501, the VHE
$\gamma$-ray variability amplitude exceeds that of the synchrotron
emission.  From this data, it is not clear whether this difference
indicates a dramatic shift in the high energy peak for Mrk 501 or that
its entire high energy emission spectrum varies more than its low
energy emission spectrum.

Correlations between the high and low energy emissions constrain the
jets of Mrk 421 and Mrk 501 through $\gamma\gamma$ transparency
estimates or the assumption that the inverse Compton (IC) mechanism
produces the high energy emission (Catanese et al.  1997).  In the
transparency arguments, a limit on the Doppler factor, $\delta$, is
derived from the requirement that the opacity due to pair production
between the $\gamma$-rays and lower energy photons be $<$1 (e.g.,
Dermer \& Gehrels 1995).  Assuming that the apparent correlation
between the VHE $\gamma$-rays and optical/UV photons observed in Mrk
421 indicates that these photons are produced in the same location,
Buckley et al. (1996) derive $\delta \ga 5$.  For Mrk 501, the
correlation between VHE $\gamma$-rays and X-rays does not lead to a
requirement for beaming through transparency arguments (Catanese et
al. 1997).

The justification for using the IC method derives from the expectation
of IC models that the X-rays and VHE $\gamma$-rays should be
correlated because they are produced by the same electrons.  Then, an
upper limit can be placed on the magnetic field, $B$, from the maximum
energy of the VHE $\gamma$-rays and the peak of the synchrotron
spectrum (cf., Buckley et al. 1998).  A lower limit on $B$ can be
derived from the assumption that synchrotron cooling dominates the
electron energy losses.  This assumption is consistent with the nearly
equal power output of the synchrotron and IC emission in these objects
and with observations of energy dependent time lags in the X-ray
emission of Mrk 421 (Takahashi et al. 1996).  By requiring that the
lower limit on $B$ is not higher than the upper limit, limits on
$\delta$ and $B$ can be derived.  For Mrk 421, assuming the VHE
spectrum extends to 5 TeV (Zweerink et al. 1997), a synchrotron
cut-off at 1 keV, a time variability scale of 1 day, and correlated
variability at 700 eV, we derive $\delta \ga 38$ and $B \ga 0.03$G.
For Mrk 501, the extent of the VHE spectrum is at least 8 TeV
(Samuelson et al. 1998), the synchrotron peak is at 100 keV, the
variability time scale is 1 day and correlated variability is observed
at 2 keV.  These inputs lead to $\delta \ga 2$ and $B \ga 0.09$G.  The
$\delta$ for Mrk 421 is quite high but it is not considerably
different than what is derived from more accurate modeling of the
emission as synchrotron self-Compton emission (Tavecchio, Maraschi \&
Ghisellini 1998).  This may indicate that Mrk 421 is beamed atypically
high or that more complicated emission mechanisms are at work in this
object.

In summary, we have evidence of differences in the multi-wavelength
emissions of Mrk 421 and Mrk 501.  Given the small sample of good
multi-wavelength campaigns on these two objects, it is not clear
whether these are genuine differences in the jets of Mrk 421 and Mrk
501, or whether they just indicate different flaring characteristics
in specific episodes.  Planned and recently completed multi-wavelength
campaigns may help resolve these issues.

\acknowledgments

This research is supported by the Department of Energy and NASA in the
US, by Forbairt in Ireland, and by PPARC in the UK.  MC acknowledges
the support of NASA grant NAG5-7070.

\end{document}